\documentclass[10pt,conference]{IEEEtran}
\usepackage{epsfig,rotating,setspace,latexsym,amsmath,epsf,amssymb,amsfonts,bm,theorem,cite,authblk, bbm,color, multirow, caption, subcaption, booktabs,  xcolor}

\IEEEoverridecommandlockouts
\allowdisplaybreaks

\title{Privacy-Preserving Spectrum Data Obfuscation in Next-Generation Spectrum Coexistence Systems \thanks{Distribution Statement A: Approved for public release: distribution unlimited. This effort was sponsored by the U.S. Government under Project Agreement NSC-25-Z514 between the National Spectrum Consortium (NSC) and the Government. The U.S. Government is authorized to reproduce and distribute reprints for Governmental purposes notwithstanding any copyright notation herein. The views and conclusions contained herein are those of the authors and should not be interpreted as necessarily representing the official policies or endorsements, either expressed or implied, of the U.S. Government.}}

\author{Yalin E. Sagduyu\IEEEauthorrefmark{1}, Tugba Erpek\IEEEauthorrefmark{1}, Sastry Kompella\IEEEauthorrefmark{1}, and Joseph A. Molnar\IEEEauthorrefmark{2}\\\IEEEauthorrefmark{1}Nexcepta Inc., Gaithersburg, MD, USA\\
  \IEEEauthorrefmark{2}Networks and Communications Systems Branch, Naval Research Laboratory, Washington, D.C., USA}

\date{}
\IEEEoverridecommandlockouts
\begin{document}
\maketitle
\thispagestyle{empty}
	
\begin{abstract}
In spectrum coexistence systems, the primary user (PU) activity must be partially disclosed for interference avoidance and coordination, while the exact spectrum occupancy information is often restricted due to security and regulatory constraints. To this end, subset-based disclosures can be employed in which the true frequency is included while ambiguity is introduced. This paper shows that naive randomized subset disclosures remain highly vulnerable to information leakage through inference, as temporal structure can be exploited to recover spectrum usage information. Under Markovian dynamics of the PU's spectrum usage, hidden Markov model (HMM) filtering with online learning is applied, while under periodic dynamics, period and phase estimations with phase-aligned statistics are used for inference. To address this vulnerability, structure-aware obfuscation strategies are introduced. For Markovian dynamics, history-based and belief-based schemes are designed to induce model mismatch and limit posterior concentration, while for periodic dynamics, cycle-dominance and decoy-based schemes are designed to manipulate phase-wise statistics and bias the inference process. These obfuscation schemes significantly hinder inference of spectrum information and offer principled disclosure solutions for spectrum coexistence systems.
\end{abstract}
\begin{IEEEkeywords}
	Spectrum obfuscation, spectrum coexistence, information disclosure, information leakage, privacy.
\end{IEEEkeywords}

\section{Introduction}
The scarcity of spectrum resources has motivated \emph{spectrum sharing} between incumbent systems (e.g., satellite and radar) and emerging (e.g., 5G and beyond) communications technologies \cite{chiriyath2017radar, labib2017coexistence, zheng2019radar, howard2022distributed, forbes20255g, forbes2025coexistent}. In \emph{spectrum coexistence} systems, some level of information from the primary user (PU), such as its temporal or spatial spectrum usage activity, must be disclosed to avoid interference. Secondary users (SUs) rely on this information to select transmit configurations, e.g., frequencies, that do not disrupt PU transmissions, making some \emph{disclosure} necessary. However, revealing exact spectrum usage is often undesirable due to security, regulatory, and operational constraints, as it may expose sensitive usage patterns for potential exploitation. This operational concern motivates spectrum  obfuscation to limit information leakage \cite{robertson2013spectrum,garnaev2017competitive,ciftcioglu2017efficient, bhattarai2017thwarting, verma2018network,dimas2019radar,grissa2021anonymous, makara2022whac, tolley2023systematization,chakraborty2024temporal}. While sharing spectrum with emerging communications systems, \emph{spectrum obfuscation} at the sensor or policy level is essential for incumbent systems to conceal their precise spectrum‑occupancy patterns and protect mission‑critical information.

In this paper, we consider the problem of \emph{spectrum obfuscation} and focus on the case of coarse, namely set-valued, information disclosed to the SUs, indicating candidate frequency bands rather than the exact occupied band. This leads to a subset-based disclosure model in which the disclosed set contains the true state with additional elements to limit leakage. While such truthful but ambiguous signaling enables spectrum coexistence, it creates an inference surface that can be further exploited. We consider two classes of spectrum access dynamics for the PU: (i) \emph{stochastic dynamics with Markovian temporal memory} (Markovian) and (ii) \emph{deterministic periodic dynamics} (periodic). In the former, spectrum occupancy evolves as a Markov process, while in the latter it follows a repeating deterministic pattern, inducing strong temporal regularity. We show that naive randomized subset disclosure, although seemingly uncertain, can be exploited to learn spectrum usage by leveraging temporal structure. While such disclosures obscure instantaneous information, they preserve statistical dependencies across time that can be systematically exploited, threatening \emph{spectrum privacy}.

Under Markovian PU activity, the SU models the system as a hidden Markov model (HMM) and applies filtering with online transition learning to infer the hidden state sequence with high accuracy. The randomized disclosures remain consistent with a fixed emission model, enabling progressive refinement of transition estimates and concentration of the posterior on the true state. Under periodic spectrum usage, the SU exploits temporal structure by estimating the period and phase offset from the disclosure sequence. Even when individual disclosures are ambiguous, temporal correlations enable phase-aligned histograms that identify dominant states within each cycle position. As a result, spectrum usage can be recovered with high reliability despite substantial instantaneous ambiguity. Overall, randomness alone does not provide effective obfuscation in the presence of temporal structure.

We show a consistent pattern across stochastic and periodic spectrum access models: \emph{temporal structure} enables \emph{accurate inference} even when individual disclosures appear ambiguous. Under randomized subset disclosure, the SU exploits Markovian dependence or periodic regularity to reconstruct spectrum usage with high accuracy. In contrast, the proposed structure-aware strategies degrade inference by disrupting the statistical cues for the SU and shape the temporal structure of disclosed information rather than relying on randomness alone.

The remainder of this paper is organized as follows. Section \ref{sec:system} formulates the spectrum obfuscation problem. Sections \ref{sec:markovian} and \ref{sec:periodic} present the obfuscation strategies for Markovian and periodic dynamics, respectively. Section \ref{sec:conclusion} concludes the paper.

\section{System Model and Problem Formulation \label{sec:system} }
\begin{figure}[t!]
	\centering
	\includegraphics[width=0.85\linewidth]{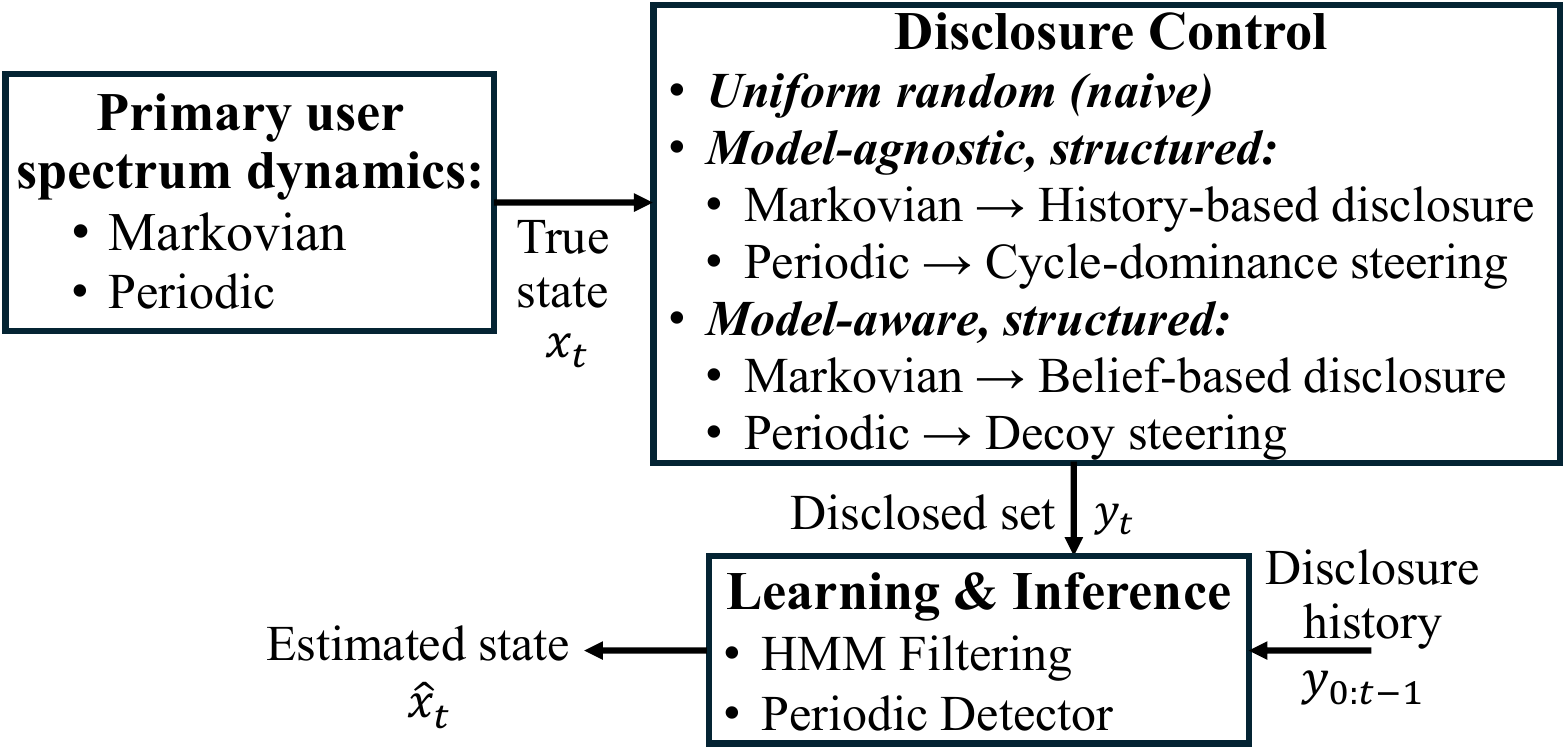}
	\caption{System model for spectrum obfuscation.}
    \vspace{-0.1cm}
	\label{fig:system_reveal}
\end{figure}

The system model for spectrum obfuscation is shown in Fig.~\ref{fig:system_reveal}. A PU occupies one of $n$ discrete frequency bands at any time $t$. At time $t$, the index of the true spectrum state is 
\begin{equation}
	x_t \in \{1,\ldots,n\},
\end{equation}
which evolves according to an underlying temporal process.
The PU observes $x_t$ and releases a subset-valued disclosure
\begin{equation}
	y_t \in \mathcal{Y} = \{ S \subseteq \{1,\ldots,n\} : |S| = m \}
\end{equation}
subject to the constraint $x_t \in y_t$. Thus, the PU does not falsify the true frequency but discloses it ambiguously through an $m$-subset. The disclosure space consists of all $\binom{n}{m}$ subsets of size $m$. For each state $i$, the set of feasible disclosures is
\begin{equation}
	\mathcal{Y}(i) = \{ y \in \mathcal{Y} : i \in y \}.
\end{equation}

The sequence $\{y_t\}_{t=1}^T$ is observed by the SU, where $T$ is the observation horizon (i.e., the total number of time steps). The SU may infer the hidden state sequence $\{x_t\}_{t=1}^T$ using a learning-based inference mechanism. The specific form of the temporal dynamics governing $\{x_t\}$ determines the SU's inference strategy and the design of the disclosure control policy. Accounting for spatial properties, the spectrum state can be extended to incorporate other transmit properties, e.g., beams and locations. In this paper, we consider two representative classes of temporal spectrum usage dynamics described below.

\textit{(i) Markovian Dynamics:} In the first class, the PU's spectrum state evolves according to a stochastic process with Markovian temporal dependence. Specifically, $\{x_t\}$ forms a first-order Markov chain with transition probabilities
\begin{equation}
	\Pr(x_t = j \mid x_{t-1} = i) = P_{ij},
\end{equation}
where $\mathbf{P} \in \mathbb{R}^{n \times n}$ is a row-stochastic transition matrix.

This model captures stochastic spectrum usage patterns in which the current state depends probabilistically on the previous state. The resulting temporal dependence can be exploited by the SU using sequential inference methods such as HMM filtering with online transition learning.

\textit{(ii) Periodic Dynamics:} In the second class, the PU's spectrum state follows a deterministic periodic pattern. There exists a sequence $\mathbf{p} = (p_1, p_2, \ldots, p_{K})$ of length $K$ such that
\begin{equation}
x_t = p_{((t-1 \bmod K)+1)}, \quad t = 1,2,\ldots,T.
\end{equation}

The sequence is repeated over time, producing temporal regularity. This deterministic structure enables the SU to estimate the period and phase offset and perform inference based on phase-aligned statistics.

\section{Spectrum Obfuscation for Markovian Source \label{sec:markovian}}

\subsection{Threat Model and HMM-Based Inference}

The SU models the system as an HMM with hidden state $x_t$ and disclosure $y_t$. The emission model is defined over the observation alphabet $\mathcal{Y}$ as
\begin{equation}
	B_{i,y} \triangleq \Pr(y_t = y \mid x_t = i), \qquad y \in \mathcal{Y},
\end{equation}
which denotes the probability of disclosing subset $y$ when the true state is $i$.  Under a nominal assumption, the SU assumes a fixed emission model in which the PU selects uniformly at random among all subsets containing the true state, yielding
\begin{equation}
	B_{i,y} =
	\begin{cases}
		\frac{1}{|\mathcal{Y}(i)|} = \frac{1}{\binom{n-1}{m-1}}, & \text{if } i \in y, \\
		0, & \text{otherwise}.
	\end{cases}
\end{equation} 
The PU may deviate from this assumed model via structured disclosure strategies, thereby inducing model mismatch at the SU. Given observations up to time $t$, the SU performs causal filtering. Let $\boldsymbol{\alpha}_t \in \mathbb{R}^n$ denote the posterior belief with components $\alpha_t(j) = \Pr(x_t = j \mid y_{1:t})$, where $y_{1:t} = (y_1,\ldots,y_t)$ is the sequence of observations up to time $t$. The belief is updated using the standard HMM filtering recursion. At time $t$, the realized disclosure is $y_t$, and the corresponding likelihood term is $B_{j,y_t}$. The posterior update is given by
\begin{equation}
\alpha_t(j) =
\frac{ B_{j,y_t} \left(\sum_{i=1}^n \alpha_{t-1}(i) P_{ij}\right) }
{ \sum_{k=1}^n B_{k,y_t} \left(\sum_{i=1}^n \alpha_{t-1}(i) P_{ik}\right) },
\end{equation}
where $i$ and $j$ index the previous and current states, respectively. The initialization is $\boldsymbol{\alpha}_0 = \boldsymbol{\pi}$, where  the initial state distribution $\boldsymbol{\pi}$ is randomly selected. The state estimate is obtained via the maximum a posteriori (MAP) rule 
\begin{equation}
\hat{x}_t = \arg\max_j \alpha_t(j).
\end{equation}

\subsection{Online Transition Learning from Partial Disclosures}

In practice, the SU does not know $\mathbf{P}$ and must estimate it online from subset disclosures. A causal learning procedure is employed that relies on soft transition estimates derived from the current belief. At time $t$, define the proxy transition matrix
\begin{equation}
	\xi_t(i,j) \propto \alpha_{t-1}(i) \hat{P}_{ij}^{(t-1)} B_{j,y_t},
\end{equation}
where $\hat{P}_{ij}^{(t-1)}$ is the current estimate of the transition probability at time $t-1$. This represents a soft estimate of the probability of transitioning from state $i$ to $j$. This proxy plays the role of transition counts in an expectation-maximization framework. The probability of switching is defined as
\begin{equation}
	w_t = 1 - \sum_{i=1}^n \xi_t(i,i),
\end{equation}
which serves as a proxy estimate of the probability that a state transition occurred at time $t$. The switching probability is updated via an exponential moving average
\begin{equation}
	\hat{\epsilon}_t = (1-\rho_t)\hat{\epsilon}_{t-1} + \rho_t w_t,
\end{equation}
where $\rho_t = \rho_0 (t+1)^{-\gamma}$ is a decaying step size. Conditioned on switching, the relative likelihood of destination states is obtained from the off-diagonal entries of $\xi_t$, yielding
\begin{equation}
	Q_{ij} = \frac{\xi_t(i,j)}{\sum_{k \neq i} \xi_t(i,k)}, \quad i \neq j,
\end{equation}
with $Q_{ii} = 0$. Each row of $\mathbf{Q}$ thus defines a probability distribution over next states given that a transition occurs.

The transition matrix estimate is reconstructed as
\begin{equation}
	\hat{\mathbf{P}}_t = (1 - \hat{\epsilon}_t) \mathbf{I} + \hat{\epsilon}_t \mathbf{Q},
\end{equation}
which provides the updated transition matrix estimate at time $t$ and separates the switching frequency from the transition destination distribution. This parameterization corresponds to a sticky Markov model, where $\hat{\epsilon}_t$ controls persistence and $\mathbf{Q}$ governs transition destinations. This separation is critical for stable online learning under partial disclosures.

\subsection{History-Based Disclosure Scheme}

The PU controls the disclosure sequence $\{y_t\}$ subject to $x_t \in y_t$, with the objective of degrading the SU's inference accuracy. This leads to a strategic information design problem in which the PU shapes the disclosure process to maximize ambiguity while maintaining correctness. The PU does not inject noise or alter the underlying signal. It selectively discloses subsets of frequencies to influence the SU's posterior belief. In the history-based scheme, the PU introduces temporal structure in the disclosure process without explicitly modeling the SU. Specifically, the PU restricts its disclosures to a circular-window codebook consisting of subsets of the form
\begin{equation}
	y = \{k, k+1, \ldots, k+m-1\} \!\!\! \mod n,
\end{equation}
yielding a structured subset family of size $n$.

At each time $t$, the PU selects a subset $y_t \in \mathcal{Y}(x_t)$ that maximizes overlap with the previously disclosed subset:
\begin{equation}
	y_t = \arg\max_{y: x_t \in y} |y \cap y_{t-1}|.
\end{equation}

This overlap-based selection rule induces temporal dependencies in the disclosure sequence that are not captured by the SU's HMM, resulting in model mismatch and degraded transition learning and filtering. Under structured subsets, the history-based scheme selects $y_t$ by evaluating overlap with the previous subset, resulting in $\mathcal{O}(nm)$ complexity per step.

\subsection{Belief-Based Disclosure Scheme}

In the belief-based scheme, the PU explicitly accounts for the SU's inference process. The PU maintains an estimate of the SU's belief $\mathbf{b}_t$, which approximates the posterior distribution over states. In practice, the PU does not have direct access to the SU's belief. Instead, the PU maintains an internal replica model of the SU's inference process and computes $\mathbf{b}_t$ using a filtering recursion based on past disclosed subsets. The parameters of this replica model, including $(\hat{\mathbf{P}}, \hat{\boldsymbol{\pi}})$, are estimated online from the disclosure history, yielding an approximate but causally consistent estimate of the SU's belief. This belief evolves according to a filtering recursion using an estimated transition model $(\hat{\mathbf{P}}, \hat{\boldsymbol{\pi}})$ learned from past disclosures.

At each time $t$, the PU selects a subset $y_t \in \mathcal{Y}(x_t)$ that minimizes the probability mass assigned to excluded states:
\begin{equation}
	y_t = \arg\min_{y: x_t \in y} \sum_{k \notin y} b_t(k).
\end{equation}
Equivalently, the PU selects subsets that retain high-probability competing states, thereby preventing the SU's belief from concentrating on the true state. This scheme directly targets the SU's posterior uncertainty and adapts to the inferred dynamics of the system. Its effectiveness depends on the accuracy of the PU's estimate of the SU's model. Under structured subsets, the belief-based scheme incurs $\mathcal{O}(n^2)$ complexity per step for belief updates and $\mathcal{O}(nm)$ for subset selection.

\subsection{Performance Evaluation}
Performance is evaluated over 300 independent realizations for statistical robustness. The SU's transition model is $
	\mathbf{P} = (1-\epsilon)\mathbf{I} + \epsilon \mathbf{Q}$,
where $\epsilon$ captures the switching probability and $\mathbf{Q}$ captures the conditional transition structure. The parameters are updated online using causal estimates of transition statistics obtained from the filtering posterior. 
For each state $i$, a switching probability $\epsilon_i \sim \mathrm{Unif}(0.003,0.01)$ and a residual-spread parameter $\eta_i \sim \mathrm{Unif}(0.01,0.05)$ are sampled. The diagonal entry is set to $P_{ii}=1-\epsilon_i$, while one randomly selected off-diagonal state is assigned probability mass $\epsilon_i(1-\eta_i)$. The remaining mass $\epsilon_i \eta_i$ is distributed across the other off-diagonal states using a Dirichlet draw. A small numerical jitter of $10^{-6}$ is added, and each row is normalized to ensure $\sum_j P_{ij}=1$, resulting in a strongly persistent Markov chain.

Given $\mathbf{P}_{\mathrm{true}}$, the corresponding stationary distribution $\boldsymbol{\pi}_{\mathrm{true}}$ is computed, and a single hidden state trajectory $\{x_t\}_{t=1}^T$ is generated. This trajectory is shared across all disclosure strategies within each realization so that performance differences arise solely from the disclosure policy. For each strategy, the transmitter generates the disclosure sequence $\{y_t\}_{t=1}^T$ based on the same underlying $\{x_t\}$, and the SU performs inference using HMM filtering with online transition learning.

\begin{figure}[t]
	\centering
	\includegraphics[width=0.94\linewidth]{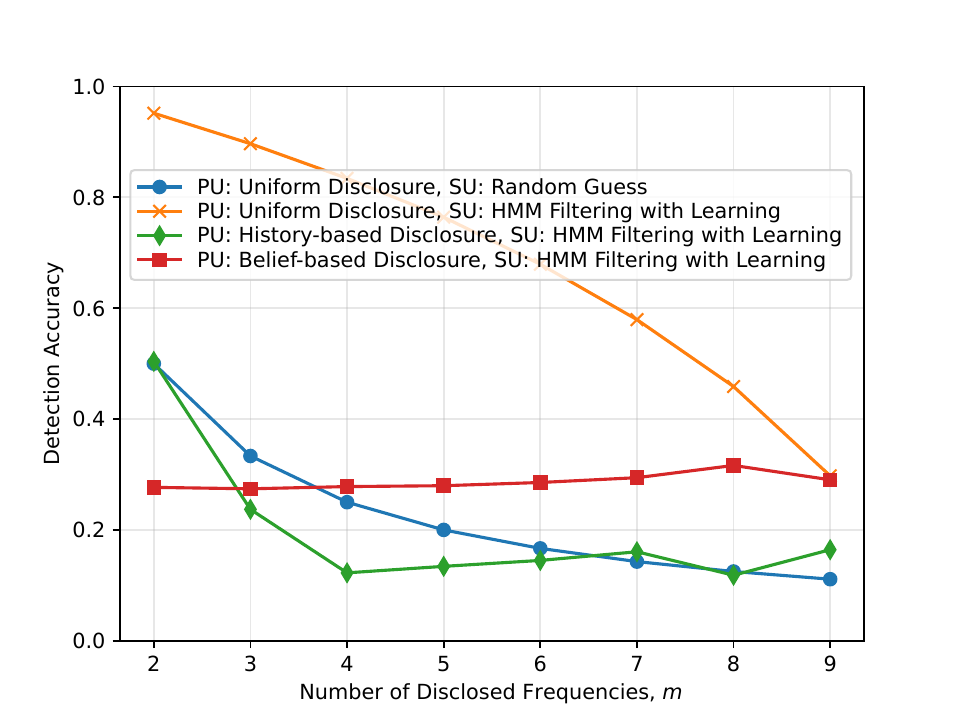}
	\caption{Detection accuracy versus number of disclosed frequencies $m$ under Markovian dynamics ($n=10$).
	}
	\label{fig:markov_m}
\end{figure}

Performance is measured in terms of detection accuracy
\begin{equation}
	\mathrm{Acc} = \frac{1}{T} \sum_{t=1}^T \mathbf{1}(\hat{x}_t = x_t),
\end{equation}
and results are averaged over independent realizations. Uniform randomized disclosure is used as the baseline. Fig.~\ref{fig:markov_m} shows the detection accuracy as a function of the number of disclosed frequencies $m$ for a fixed total number of bands $n=10$. Under uniform randomized disclosures, the SU achieves high detection accuracy across all values of $m$ (particularly for smaller $m$), significantly outperforming random guessing (which is successful with probability $1/m$). This shows that, despite substantial instantaneous ambiguity, the temporal dependence of the Markov process provides sufficient structure for accurate inference through sequential filtering and model adaptation. 

These obfuscation schemes substantially reduce detection performance. The history-based disclosure introduces temporal correlations in the disclosure sequence that violate the SU's assumed emission model, thereby degrading filtering consistency over time. The belief-based disclosure further suppresses the inference performance by shaping the SU's posterior distribution to retain high-probability competing states, preventing concentration on the true state and limiting the effectiveness of sequential updates. The performance of obfuscation methods depends on the observation horizon $T$. For example, for $n\!=\!10$ and $m\!=4$, $T\!=\!500$ suffices for the belief-based scheme to achieve detection accuracy within 5\% of that at $T\!=\!10000$. However, the history-based scheme requires $T\!=\!4000$ to reach the 5\% gap of detection accuracy at $T\!=\!10000$.

\begin{figure}[t]
	\centering
	\includegraphics[width=0.94\linewidth]{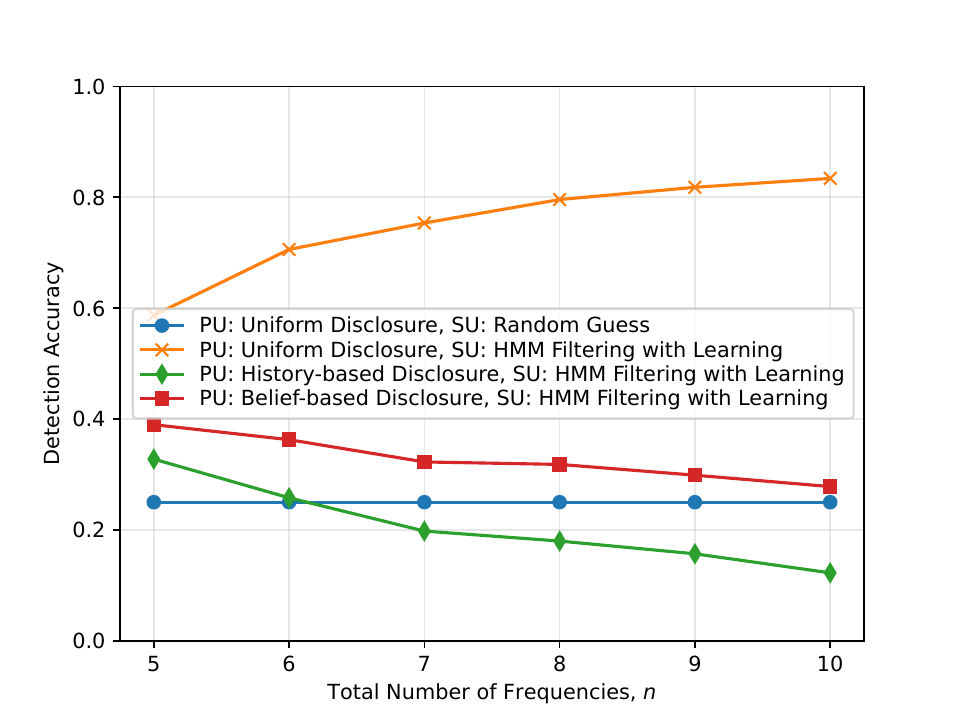}
	\caption{Detection accuracy versus total number of frequencies $n$ under Markovian dynamics ($m=4$). 
	}
	\label{fig:markov_n}
\end{figure}

Fig.~\ref{fig:markov_n}  shows detection accuracy as a function of the total number of frequencies $n$ for a fixed subset size $m=4$. As $n$ increases, detection accuracy of the SU increases and the gain over random guessing becomes more pronounced. Detection accuracy for the proposed obfuscation schemes is much lower due to increased ambiguity. The belief-based scheme consistently achieves the lowest detection accuracy, indicating its robustness in maintaining uncertainty even as the state space grows. Overall, these results show that (i) naive randomized subset disclosure remains highly vulnerable under Markovian dynamics, and (ii) structure-aware disclosure strategies significantly degrade the inference process by disrupting model consistency and limiting posterior concentration.
 
\section{Spectrum Obfuscation for Periodic Source \label{sec:periodic}}

\subsection{Periodic Mode Detection}

The SU exploits periodic structure by estimating both the period $K$ and the phase offset $\phi$ from the observed disclosure sequence. Let $T_1 \le T$ denote the number of observations used for period and phase estimation. The estimation is performed using only the first $T_1$ samples of the same disclosure stream, after which the estimates are frozen and used for online decoding. The estimation may be performed over an initial observation window or jointly with inference over the full sequence. In the numerical results, we set $T_1 = T$, i.e., parameter estimation and inference use the same observation horizon. To estimate the period, the SU maximizes the average Jaccard similarity between subsets separated by lag $K$:
\begin{equation}
	\hat{K} = \arg\max_{K \in [K_{\min},K_{\max}]} 
	\frac{1}{T_1-K} \sum_{t=K}^{T_1} 
	\frac{|y_t \cap y_{t-K}|}{|y_t \cup y_{t-K}|}.
\end{equation}
In case of multiple maximizers (e.g., due to periodic multiples), the smallest $K$ is selected. Given $\hat{K}$, the phase offset $\phi$ is estimated by selecting the offset that maximizes the peakedness of per-phase histograms. Let
\begin{equation}
	C_r(i) = \sum_{t=1}^{T_1} \mathbf{1}\{i \in y_t\} \cdot \mathbf{1}\{(t+\phi)\bmod K = r\}
\end{equation}
denote the count of how often state $i$ appears in phase $r$. The offset is chosen as
\begin{equation}
\hat{\phi} = \arg\max_{\phi \in \{0,\ldots,\hat{K}-1\}} 
\sum_{r=0}^{\hat{K}-1} \left( C_r(i_r^\ast) - \max_{j \neq i_r^\ast} C_r(j) \right),
\end{equation}
where
\begin{equation}
i_r^\ast = \arg\max_i C_r(i).
\end{equation}

This objective favors phase alignments that produce sharply peaked histograms, thereby enabling reliable identification of a dominant state within each cycle position.

After estimating $(\hat{K}, \hat{\phi})$, the SU maintains phase-wise counts online and decodes the state at time $t$ as
\begin{equation}
	\hat{x}_t = \arg\max_i C_r(i), \quad r = (t+\hat{\phi}) \bmod \hat{K}.
\end{equation}

This periodic detector is highly effective and can achieve near-perfect recovery under nominal (random) disclosures. 

\subsection{Spectrum Obfuscation as Periodic Structure Manipulation}

Obfuscation is achieved by disrupting the phase-wise statistics that the SU relies on. Since the SU's decoding is based on identifying dominant states within each cycle position, the PU seeks to flatten or manipulate these histograms. The PU maintains its own internal estimate of the phase counts, $\tilde{C}_r(i)$, which approximates the SU's counts, and uses this to guide disclosure decisions. These estimates are computed using an internal replica of the SU's periodic detector based solely on past disclosures, without access to SU's true internal state. Specifically, the estimates are updated online using only past disclosed subsets, ensuring that the PU operates causally. The PU assumes knowledge of the nominal period $K$ and uses a fixed internal phase reference, whereas the SU estimates $(K,\phi)$ from an initial segment of the same disclosure stream.

\subsection{Cycle-Dominance Steering (History-Based)}

In the first scheme, the PU does not explicitly model the SU's estimation process but instead manipulates the phase-wise dominance structure. At time $t$, let $r = (t \bmod K)$ denote the cycle position. The PU selects a subset $y_t \in \mathcal{Y}(x_t)$ that minimizes the dominance of the true state within that phase. Define the post-update dominance margin as
\begin{equation}
	\Delta(y) = \left( \tilde{C}_r(x_t) + 1 \right) - \max_{j \in y, j \neq x_t} \left( \tilde{C}_r(j) + 1 \right).
\end{equation}
The PU selects
\begin{equation}
	y_t = \arg\min_{y: x_t \in y} \Delta(y),
\end{equation}
thereby reducing the gap between the true state and competing states. The PU promotes diversity by increasing the aggregate counts of competing states and enforces temporal smoothness by favoring subsets that overlap with previously disclosed subsets. This results in flatter histograms across phases, making the SU's decision unreliable. When multiple subsets achieve the same objective value, deterministic tie-breaking is used. The cycle-dominance scheme maintains phase counts with $\mathcal{O}(nK)$ memory and $\mathcal{O}(m)$ update cost per step, while subset selection requires $\mathcal{O}(nm)$.

\subsection{Belief-Based Decoy Steering}

In the second scheme, the PU explicitly targets the SU's periodic detector by steering its decisions toward a specific incorrect state. The PU operates fully causally and does not rely on ground-truth phase labels or offline rehearsal. No separate training trajectory or oracle knowledge of phase indices is used; all decisions are based solely on the evolving estimates denoted by $\tilde{C}_r(i)$. At time $t$, the PU identifies a decoy state
\begin{equation}
	d_t = \arg\max_{j \neq x_t} \tilde{C}_r(j),
\end{equation}
which is the most dominant competing state in the current phase. The PU then selects a subset $y_t \in \mathcal{Y}(x_t)$ that includes both the true state and, when feasible, the decoy state.

Among such subsets, the PU minimizes the dominance margin while promoting decoy inclusion. This can be expressed as a lexicographic objective:
\begin{equation}
	y_t = \arg\min_{y: x_t \in y} \bigg( \Delta(y), -\mathbf{1}\{d_t \in y\}, -\sum_{j \in y \setminus \{x_t\}} \tilde{C}_r(j) \bigg).
\end{equation}

This hierarchical objective prioritizes reducing true-state dominance, then maximizing decoy inclusion, and finally increasing the aggregate strength of competing states, consistent with the selection rule. This scheme systematically pushes the SU's estimate toward the decoy state while suppressing the true state's dominance, effectively inducing structured misclassification. The belief-based causal decoy scheme maintains phase counts with $\mathcal{O}(nK)$ memory and $\mathcal{O}(m)$ update cost per step, and requires $\mathcal{O}(n)$ for decoy selection, yielding $\mathcal{O}(nm)$ overall.

\subsection{Performance Evaluation}

Performance is evaluated over $300$ independent realizations. In each realization, a periodic pattern of length $K=100$ is generated by sampling entries from $\{1,\ldots,n\}$ such that when $K \geq n$, each state appears at least once in the period, while for $K < n$, all entries are distinct. The state sequence is constructed according to the periodic model, and the same trajectory is used across all disclosure strategies within each realization. Detection accuracy is computed as in the Markovian case and averaged over all realizations. The SU employs a periodic mode detector that estimates the underlying period and phase offset and reconstructs the state sequence using phase-aligned statistics. 

\begin{figure}[t]
\vspace{-0.4cm}
	\centering
	\includegraphics[width=0.94\linewidth]{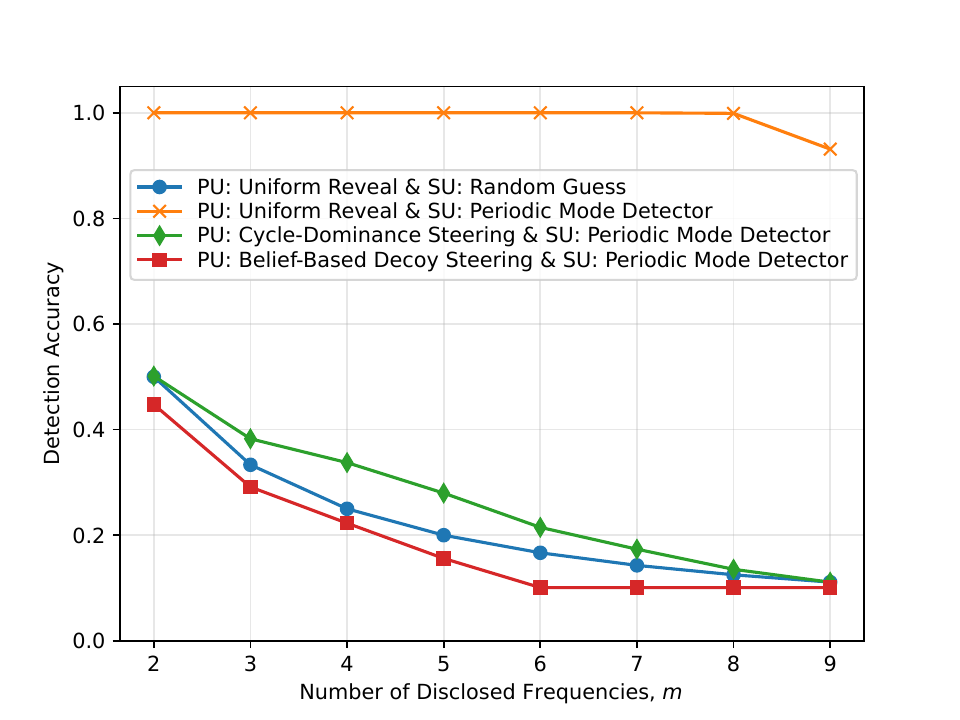}
	\caption{Detection accuracy versus number of disclosed frequencies $m$ under periodic dynamics ($n=10$). 
	}
	\label{fig:periodic_m}
	\centering
	\includegraphics[width=0.94\linewidth]{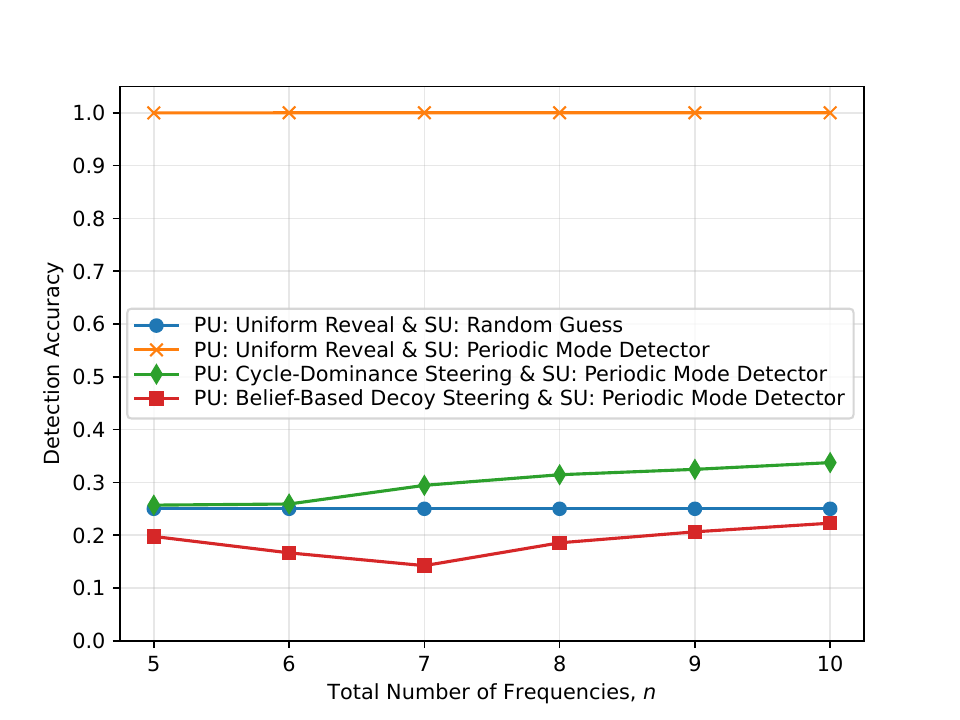}
	\caption{Detection accuracy versus total number of frequencies $n$ under periodic dynamics ($m=4$). 
	}
	\label{fig:periodic_n}
\end{figure}

Fig.~\ref{fig:periodic_m} shows detection accuracy as a function of the number of disclosed frequencies $m$ for $n=10$. Under uniform randomized disclosures, the SU achieves near-perfect detection accuracy across a wide range of $m$, demonstrating that periodic temporal structure can be reliably exploited even in the presence of significant instantaneous ambiguity. 

These obfuscation schemes substantially reduce detection performance. The detection accuracy decreases as $n$ increases. The cycle-dominance steering scheme flattens phase-wise histograms, reducing the dominance of the true state within each cycle position and weakening the SU's phase-based decision rule. The belief-based decoy steering scheme further degrades performance by systematically promoting competing states, causing the SU to converge toward incorrect phase-wise estimates and introducing persistent bias in the inferred sequence.

Fig.~\ref{fig:periodic_n} shows detection accuracy as a function of the total number of frequencies $n$ for a fixed subset size $m=4$. As $n$ increases, the SU's performance under uniform disclosures remains high, highlighting the robustness of periodic inference. In contrast, both proposed schemes continue to significantly reduce detection accuracy, with the decoy-based approach achieving the largest degradation and maintaining strong suppression of inference accuracy even as the state space grows.  The performance improves with observation horizon $T$. For example, when $n=10$ and $m=4$, $T=700$ suffices to achieve detection accuracy within 5\% of that at $T=10000$ for all obfuscation schemes considered. These results show that periodic behavior creates a strong inference advantage under naive randomized disclosure, but also provides a structured opportunity for targeted obfuscation through manipulation of phase-wise statistics and inference decision biases.

\section{Conclusion \label{sec:conclusion}}

This paper studied spectrum obfuscation under subset-based disclosure in coexistence systems, where partial information is shared without revealing exact spectrum usage. We showed that disclosure-level ambiguity is insufficient when the underlying process exhibits temporal structure. In both stochastic (Markovian) and periodic settings, temporal dependencies can be exploited to reconstruct spectrum occupancy even when individual disclosures seem to be highly uncertain. To address this privacy risk, we introduced structure-aware obfuscation strategies that adapt to the spectrum dynamics and inference process. For Markovian dynamics, these strategies disrupt model consistency and limit sequential refinement, while for periodic dynamics they weaken phase-wise dominance and introduce systematic biases in inference decisions. This broader design principle preserves the secrecy of spectrum occupancy through structured disclosure, and can be integrated with other spatial properties, sensor and policy-level obfuscation strategies, and commercial impact studies. Overall, this obfuscation approach significantly reduces the accuracy of inferring spectrum usage information of incumbent systems without relying solely on randomized ambiguity, directly supporting emerging spectrum coexistence operations.

\bibliographystyle{IEEEtran}
\bibliography{references_reveal}

\end{document}